# Planar hole-doping concentration and effective three-dimensional hole-doping concentration for single-layer high-$T_c$ superconductors

Tatsuya Honma[a,b] and Pei Herng Hor[a]

[a] Texas Center for Superconductivity and Dept. of Physics, University of Houston, Houston, TX. 77204-5002, USA.

[b] Dept. of Physics, Asahikawa Medical College, Asahikawa, Hokkaido 078-8510, Japan.

## Abstract

We propose that physical properties for the high temperature superconductors can be addressed by either a two-dimensional planar hole-doping concentration ($P_{pl}$) or an effective three-dimensional hole-doping concentration ($P_{3D}$). We find that superconducting transition temperature ($T_c$) exhibits a universal dome-shaped behavior in the $T_c$ vs. $P_{3D}$ plot with a universal optimal doping concentration at $P_{3D} \sim 1.6$ x $10^{21}$ cm$^{-3}$ for the single-layer high temperature superconductors.

Key-wards; Room-temperature thermoelectric power ; hole-doping concentration ;  Hall number ; superconducting transition temperature

## 1. Introduction

In high temperature superconductors (HTS) hole content per CuO$_2$ plane ($P_{pl}$) can be directly determined from the content of the cation dopant in the pure cation doped La$_{2-x}$Sr$_x$CuO$_4$ (SrD-La214) and Y$_{1-x}$Ca$_x$Ba$_2$Cu$_3$O$_6$ (CaD-Y1236). Most recently, based on the thermoelectric power at room temperature ($S^{290}$) of the SrD-La214 and CaD-Y1236, a universal $S^{290}(P_{pl})$-scale (hereafter $P_{pl}$-scale) [1] is constructed as new scale in contrast to $T_c(P_{Tc})$-scale (hereafter $P_{Tc}$-scale) which was defined by a relation of $T_c/T_c^{max} = 1 - 82.6(P_{Tc} - 0.16)^2$. While the $P_{Tc}$ is intrinsically equal to $P_{pl}$ in SrD-La214 [2], it is different in other systems. Using the $P_{pl}$-scale, the maximum in $T_c$ ($T_c^{max}$) was no longer universally pinned at $P_{pl} = 0.16$, it depended on the specific material system of HTS. However, many experimental data were interpreted using the $P_{Tc}$-scale by taking $P_{Tc} = P_{pl}$ [3].

In-plane Hall number ($n_H = 1/eR_H$), where $R_H$ is in-plane Hall coefficient and $|e|$ is electron charge, has physical meaning of the mobile carrier concentration per volume and is a three-dimensional (3D) quantity. But, the $P_{pl}$ is intrinsically a two-dimensional (2D) quantity. Since both concentrations monitor doped carriers, the proper extension of $P_{pl}$ is expected to be comparable to $n_H$. When the planar carriers exist in the block layer with one CuO$_2$ plane, we can define an effective 3D hole-doping concentration ($P_{3D}$) in terms of $P_{pl}$ by a relation of $P_{3D} \equiv P_{pl} \times (N_l/V_{u.c.})$. Here, $V_{u.c.}$ and $N_l$ are the unit cell volume and the number of CuO$_2$ plane per unit cell, respectively. Since $P_{3D}$ is defined on the universal 2D $P_{pl}$-scale, this definition has qualitatively taken into account the charge de-confinement effect of the holes in cuprates. Therefore $P_{3D}$ can be viewed as the "**effective**" 3D hole-doping concentration even when holes are completely confined in CuO$_2$ planes.

In this paper we make a clear distinction between $P_{pl}$ and $P_{3D}$. We show that the present $P_{3D}$ is comparable with $n_H$ and that the $T_c/T_c^{max}$ vs. $P_{3D}$ exhibits a universal dome-shaped curve with the universal optimal hole-doping concentration $P_{3D}^{opt.} = 1.6$ x $10^{21}$ cm$^{-3}$ for single-layer HTS. We find that the $P_{Tc}$-scale is identical to the $P_{3D}$-scale. The detail is reported in Ref. 1 and 5.

## 2. Experimental

The analyzed data are collected from the literatures [4,6-15] whenever the $P_{pl}$ can be reliably determined by $P_{pl}$-scale. For the calculation of $P_{3D}$, we used the typical value of the unit cell volume [5].

## 3. Results and discussion

Figure 1 shows the $n_H$ as a function of $P_{3D}$ for the single-layer SrD-La214, OD-Hg1201, OD-Tl2201 and CD-Bi2201. The plotted $n_H$ come from the polycrystalline samples for SrD-La214 [12,13] and OD-Tl2201 [4,14,16] and the single crystals for SrD-La214 [10-12] and CD-Bi2201 [7-8]. In the SrD-La214 and OD-Tl2201, the $R_H$ of the polycrystalline samples is experimentally confirmed to be almost equal to the in-plane $R_H$ of the single crystals [12,17]. There are three linear $n_H(P_{3D})$ regimes (regime-I, II and III). In regime-I for $P_{3D} \leq 5.5 \times 10^{20}$ cm$^{-3}$, $n_H$ is identical to $P_{3D}$. At $P_{3D} = 5.5 \times 10^{20}$ cm$^{-3}$, the slope of linear $n_H(P_{3D})$ suddenly changes from 1 to ~3.2. In the regime-III for $P_{3D} \geq 1.6 \times 10^{21}$ cm$^{-3}$, the linear $n_H(P_{3D})$ changes slope to 25. The observed rapid increase in $R_H$ may relate to the change in sign of $R_H$ observed in the overdoped SrD-La214 [12]. We need to emphasize that this systematic behaviour for the single-layer HTS is not governed by the $P_{pl}$, but by the $P_{3D}$. In the inset of fig.1, we plot the same data set of $n_H$ as a function of $P_{pl}$. The $n_H$ for CD-Bi2201 and OD-Tl2201 do not follow that of SrD-La214, and the three physically distinct regimes can not be resolved.

Figure 2 shows $T_c$ as a function of $P_{3D}$ for SrD-La214 [6,15], OD-Hg1201 [9] and CD-Bi2201 [7,8]. The superconductivity appears at ~$5.5 \times 10^{20}$ cm$^{-3}$ where is corresponding to the boundary between the regime-I and -II. The $T_c^{max}$ universally appears at ~$1.6 \times 10^{21}$ cm$^{-3}$ where is corresponding to the boundary between regime-II and -III. The inset shows the $T_c/T_c^{max}$ vs. $P_{3D}$ of the same data set. The $T_c/T_c^{max}$ for SrD-La214, OD-Hg1201 and CD-Bi2201 follow the same dome-shaped curve. Now we can pin down the absolute 3D optimal hole-doping concentration in a relation of $T_c/T_c^{max} = 1 - 83.64(P_{3D} \times 10^{-22} - 0.159)^2$. It is clear that the $P_{Tc}$-scale is not planar hole-doping concentration but physically identical to our defined $P_{3D}$. Therefore, we can understand why the $P_{Tc}$-scale worked in the earlier doping-dependence of $T_c$ studies [3]. However, we need to emphasize that the $P_{Tc}$-scale should be interpreted in the contexts of $P_{3D}$ as the proper carrier scale for 3D "bulk" cuprate properties.

In summary, we have shown that for HTS there are two types of hole-doping concentration depending on the dimensionality, that is, $P_{3D}$ and $P_{pl}$. Combining these two, we have a complete working scale to address various physical properties for all HTS. Indeed, we see that $n_H$ and the magnitude of $T_c$ are governed by $P_{3D}$, while pseudogap physics were described by $P_{pl}$ [1].

### Acknowledgments

One of us (T.H.) would like to thank Dr. M. Tanimoto of Asahikawa Medical College for providing the administrative convincement for this study. This work was supported by the State of Texas through the Texas Center for Superconductivity at the University of Houston.

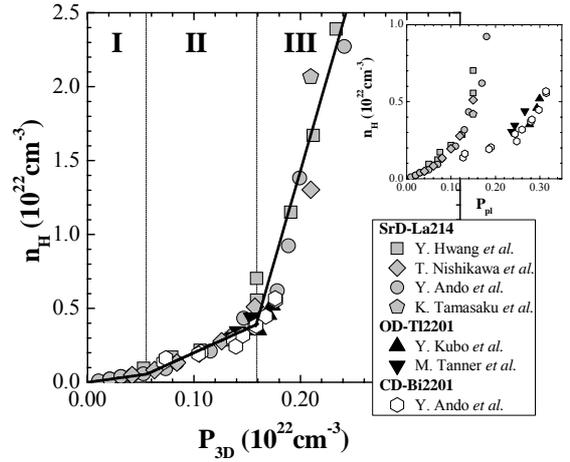

Figure 1 $n_H$ vs. $P_{3D}$ for the single-layer HTS. The inset shows $n_H$ vs. $P_{pl}$.

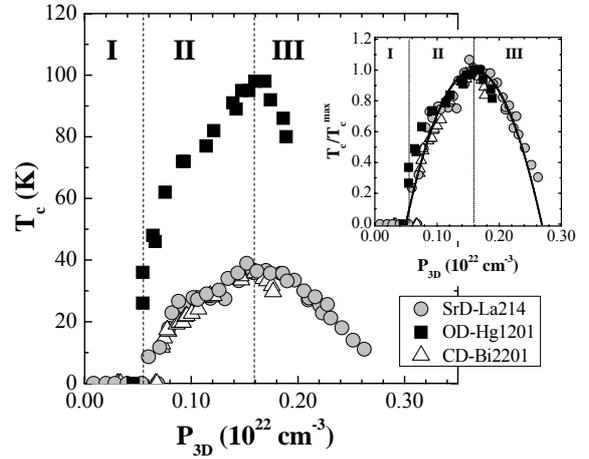

Figure 2 $T_c$ vs. $P_{3D}$ for the single-layer HTS. The inset shows $T_c/T_c^{max}$ vs. $P_{3D}$.

### References

[1] T. Honma *et al.*, Phys. Rev. B **70** (2004) 214517.

[2] M.R. Presland *et al.*, Physica C **176** (1991) 95.

[3] For example, J.L. Tallon, J.W. Loram, Physica C **349** (2001) 53.

[4] S.D. Obertelli, J.R. Cooper, J.L. Tallon, Phys. Rev. B **46** (1992) 14928.

[5] T. Honma, P. Hor, Supercond. Sci. Tech. **19** (2006) 907.

[6] P.G. Radaelli *et al.*, Phys. Rev. B **49** (1994) 4163.

[7] Y. Ando *et al.*, *Phys. Rev. B*, **61** (2000) R14956.

[8] Y. Ando, T. Murayama, S. Ono, Physica C **341-348** (2000) 1913.

[9] A. Yamamoto, W. Hu, S. Tajima., Phys. Rev. B **63** (2000) 024504.

[10] Y. Ando, Phys. Rev. Lett. **87** (2001) 017001 ; Phys. Rev. Lett. **92** (2004) 197001

[11] K. Tamasaku *et al.*, Phys. Rev. Lett. **72** (1994) 3088.

[12] H.Y. Hwang *et al.*, Phys. Rev. Lett. **72** (1994) 2636.

[13] T. Nishikawa, J. Takeda, M. Sato, J. Phys. Soc. Jpn. **63** (1995) 1441.

[14] M.A. Tannar *et al.*, Physica C **185-189** (1991) 1247.

[15] S. Komiya *et al.*, Phys. Rev. Lett. **94** (2005) 207004.

[16] Y. Kubo *et al.*, Phys. Rev. B **43** (1991) 7875.

[17] T. Manako, Y. Kubo, Y. Shimakawa, Phys. Rev. B **46** (1992) 11019.